# RFID Security Using Lightweight Mutual Authentication and Ownership Transfer Protocol

Amol Bandal[1] and Shankar Nawale[2]

[1]Department of Computer Engineering, Sinhgad Institute, Lonavala, Maharashtra, India
amolbd1987@gmail.com
[2]Department of Electronics & Telecommunication, Sinhgad Institute, Lonavala, Maharashtra, India
shankarnawale125@gmail.com

## ABSTRACT

*In recent years, radio frequency identification technology has moved into the mainstream applications that help to speed up handling of manufactured goods and materials. RFID tags are divided into two classes: active and passive. Active tag requires a power source that's why its cost is more than passive tags. However, the low-cost RFID tags are facing new challenges to security and privacy. Some solutions utilize expensive cryptographic primitives such as hash or encryption functions, and some lightweight approaches have been reported to be not secure.*

*This paper describes a lightweight Mutual authentication and ownership transfer protocol utilizing minimalistic cryptography using Physically Unclonable Functions (PUF) and Linear Feedback Shift Registers (LFSR). PUFs and LFSRs are very efficient in hardware and particularly suitable for the low-cost RFID tags. To functioning security in low cost RFID tag minimum gate requirement is 2000 gates. To implement security protocols using PUF and LFSR functions need only approx 800 gates. In this paper it is explained how we can authenticate and transfer ownership of low cost RFID tag securely using LFSR and PUF as compared to existing solutions based on hash functions.*

## 1. INTRODUCTION

In recent years, there has been a notable increase in the deployment of RFID (Radio frequency Identification) tags in supply chain management, smart appliances and access control applications. The standardization bodies such as the EPC global and GSI (Global Standardization International) are working together to ensure the proliferation of RFID applications by proposing and managing a global standard for RFID tags. However, the main challenge to the proliferation of RFID technology is to provide security solutions for low-cost passive RFID tags. Main reason of not using RFID tag at all places is the cost. As compare to barcode RFID tags having more cost and to reduce the cost of tag there is need to design RFID tag with minimum hardware requirement. In case of passive tags with minimum hardware requirement there is main challenge to provide security.

Radio frequency identification (RFID) chips are used everywhere. A number of examples can be quoted where RFID technology has been implemented—companies and laboratories use them as access keys, to start their cars, and as inventory tracking devices. Drug manufacturers rely on chips to track pharmaceuticals. For protection of information, RFID signals can be encrypted using suitable algorithms. RFID tags used in different applications like passport are encrypted so that data like person name, age and other sensitive data remain protected. But then, most of the commercial RFID tags do not include security as it is very expensive.

There are different types of RFID exist, in general we can divide RFID devices into two classes: active and passive [2].The tags work by broadcasting a few bits of information to specialized electronic readers. Most commercial RFID tags are passive emitters and have no onboard battery: these tags get activated by the reader power. Once activated, these chips broadcast their signal indiscriminately within a certain range, usually a few inches to a few feet. However, active RFID tags with internal power can send signals to hundreds of feet. Active tags have more cost than the passive tags as they require power source.

| Reserve Memory contains Kill & Access password |
| --- |
| TID (Identification of tag capability) |
| EPC (Electronic product code) |
| User Memory (user specific data) |

Figure 1. Memory map of Gen2 tag

The latest EPC Tag Data standards for tags (also known as EPC-C1G2 tags and Gen2 tags) include a complete specification for the 96 bit EPC (Code) [7]. In the specification, the tag memory is divided into four distinct banks, each of which may compromise one or more memory words, where each word is 16 bits long. These memory banks are described as "Reserve", "EPC", "TID" and "User" [Fig.1] The "Reserve" memory banks contains kill and access passwords; the "EPC" memory bank contains electronic product code information (EPC) which is used for identifying the object to which the tag will be attached; the "TID" memory bank contains data that reader to identify the tag's capability; and the "User" memory bank is intended to contain user-specific data. Reserve memory bank is read and write protected. The role of the 32 bit kill password is to permanently deactivate the tag. The 32 bit access password is used for locking any memory bank of a Gen2 tag.

| Header (H) 8 bit | EPC Manager (EM) 28 bit | Object Class (OC) 24 bit | Serial Number (SN) 36 bit |
| --- | --- | --- | --- |

Figure 2. Structure of 96 bit EPC

In addition, the data stored in the EPC Gen2 tag has four fields [7]: the 8-bit Header, H(A fixed identifier which guarantees the unique namespace), the 28-bit EPC Manager, EM (which represents company or an organization); the 24-bit object class (OC represents unique type of product within company); the 36-bit Serial Number, SN (which represents unique item within a particular product). Figure 2 shows the EPC of Gen2 tag.

RFID system mainly consists of three components which are Tag, Reader and database, as shown in Figure 3. Reader is also called Transceiver and Tag is also called Transponder. Reader launches a specific radio frequency of wireless waves to Tag. When Tag sends its internal data after receiving them, Reader will be in sequence to receive data and authentication Tag . When label is into the magnetic field, it will receive the radio frequency signals emitted by Reader, with the energy by virtue of induced current to send out the product information stored in the chip (Passive Tag), or to take the initiative to send a frequency signal (Active Tag); reader reads the information, decodes it and sends it to a central information system to run the related data processing.

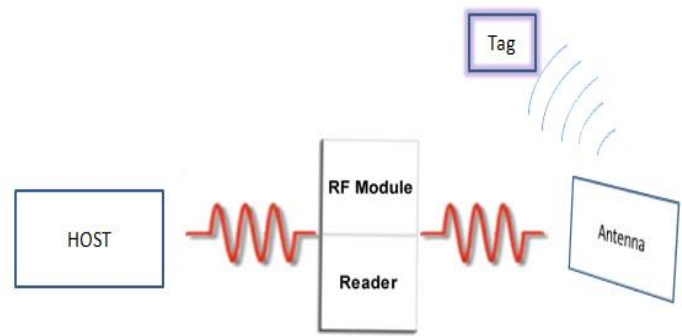

Figure 3. Working of RFID

### 1.1. Motivation:

Despite the increasing popularity of RFID technology, the electronic information it deals with, may not be as secure as was once thought.

Problems with RFID
1. RFID system can be easily disrupted.
2. RFID reader Collision/Interference
3. RFID Tags Can Be Read without Your Knowledge

Specifically, the pressing issues in RFID security [3],[4],[11] include eavesdropping and tag cloning. Eavesdropping can lead to loss of privacy, since a tag may reveal its identity to any reader within its range. Cloning means that the attacker can use the eavesdropped information to access either the tag or a legitimate reader or unauthorized access privileges. As RFID tags are placed in heavily targeted items such as credit cards and passports, it is clear that RFID needs security measures that are both affordable and reliable.

In case of RFID system reader reads the information from the tag wirelessly. Wireless link between reader and tag is another entry point for attacker to attack the RFID system. Once additional systems are compromised, all types of adverse consequences to the IT infrastructure are possible, including loss of confidentiality, integrity, and availability.

As in day to day life use of RFID tag is increasing in applications like passports, credit cards etc. we need have RFID tag with less hardware. We can't have batteries and large circuit on passport or credit card. As less hardware cost will be less and we can use it in many applications. Passive tags without batteries don't have processing ability. Passive tags with no batteries have maximum 2000 gates available for security measures [10]. We can't go for encryption techniques like AES, MD5 or SHA because to implement all this required gate count is more than 3000. [8]

## 2. LITERATURE REVIEW

There have been several hash based solutions that create authentication for RFID systems. Solutions relying on encryption or hash functions become prohibitively expensive in the case of many low-cost RFID systems.

### 2.1. Lightweight Solutions for Mutual Authentication:

#### 2.1.1. LMAP (Lightweight Mutual Authentication Protocol)

Low-cost Radio Frequency Identification (RFID) tags are devices of very limited computational capabilities, where only 250-3000 logic gates can be devoted to security-related tasks. Many proposals have recently appeared, but all of them are based on RFID tags using classical cryptographic primitives such as PRNGs, hash functions, block ciphers, etc. LMAP protocol [5] need only 300 gates. It uses 96 bit length index pseudonym as index of table where information of tag is stored .It uses only bitwise XOR, bitwise OR ( V ) , bitwise AND (^) operations.

### 2.1.2. Hopper and Blum (HB) Protocol :

Hopper and Blum present [6] an authentication scheme called the HB protocol based on the LPN (Learning Parity with Noise) problem. The scheme uses only XOR and requires only two moves. However, the HB protocol is only secure against passive eavesdroppers, and it is not secure against an active adversary with the ability to query tags.

HB protocol relies on the computational hardness of Learning Parity with Noise (LPN) problem, and not on classical symmetric key cryptography solutions.

### 2.1.3. Juels and Weis (HB+) Protocol:

Juels and Weis present an extended authentication scheme to the HB protocol, called the HB+ protocol[6], that is secure against an active adversary. The scheme uses only the XOR calculation and requires three moves. Juels and Weis modified the HB protocol and showed the modified protocol (HB+) to be secure against active attacks.

Although Juels and Weis showed HB+ to be secure against active attacks, Gilbert showed that HB+ is not secure against a simple man-in-the-middle attack that was not considered in former.

### 2.2. RFID protocols using Hash Function:

There are different hash functions [8] available to perform encrypt operations like SHA-1, SHA-256, AES, MD4 , MD5 etc. problems with all this algorithms is the computation overhead and power requirement for total number of operations. Most of the cryptographic techniques mentioned are applicable for Active tags as they requires more computing.

The Gate count available for Security of Passive RFID tag is only 2000 gates [10]. As mentioned in table to implement SHA requirement is more than 8000 gates [8]. In case of MD4, MD5 and AES gate requirement is more than the available peak value ie 2000. In this paper whatever algorithm is explained is based on LFSR and PUF function. As per the referred paper [1] this scheme demands only 784 gates out of available 2000 gates.

Table 1. Comparison of Hash Function

| Algorithm | Chip Area Gate Equivalent |
|---|---|
| SHA-256 | 10,868 Gates |
| SHA – 1 | 8120 Gates |
| MD5 | 8400 Gates |
| MD4 | 7350 Gates |
| AES | 3400 Gates |

# 3. SYSTEM ARCHITECTURE AND DETAILED DESIGN

## 3.1. Mutual Authentication protocol

To achieve mutual authentication between reader and tag, tag need to verify that received messages are generated by the authenticated reader and reader need to verify received messages are generated by the tag that the reader wants to authenticate.

### 3.1.1. Setup Phase:

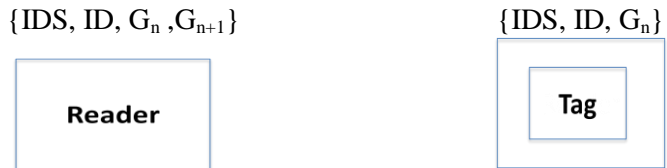

Figure 4. Set up phase

In this phase, both the reader and the tag are preloaded with a tuple of three secrets {IDS, ID, $G_n$} that is known only by themselves, where IDS is the index to tag's ID and is updated in each round, while $G_n$ is a greeting number. Since the reader needs to authenticate multiple tags, it maintains a table that stores the tuples for all tags that it wants to authenticate. Besides, the reader and the tag share a random permutation function F mapping within range [1, q], where log q is the bit length of IDS. LFSR can be used as the function F and is implemented using Linear Feedback Shift Register technique. Reader also stores $G_{n+1}$ as another greeting for each tag.

Tag implements function P which is random permutation function mapping within range [1, q]. P is implemented based on physically unclonable function (PUF) and its output is used to verify tag's identity.

### 3.1.2. Authentication Phase:

To achieve mutual authentication between a reader and a tag, the following two properties should be satisfied.

Property 1: The tag can verify that the received messages are generated by the reader that the tag intends to authenticate.

Property 2:
The reader can verify that the received messages are generated by the tag that the reader intends to authenticate.

Step 1: The reader continuously broadcasts Req message to request for any tags.

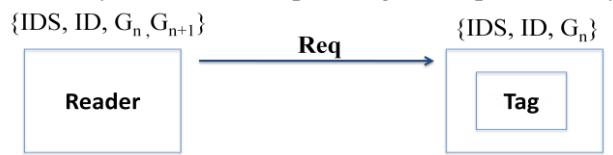

Step 2: Receiving Req from the reader, the tag responds with its IDS, which does not reveal the tag's ID.

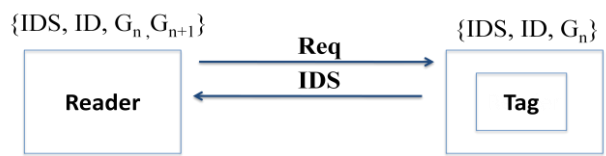

Step 3: The reader uses IDS to look up the corresponding greeting $G_n$ for this tag, then it sends back ID$\oplus G_n$ to the tag. In this response, ID is protected by $G_n$. As long as $G_n$ is not revealed, tracking the tag by its ID will be impossible. Receiving ID$\oplus G_n$, the tag verifies the correctness of this response using its own ID and $G_n$. If correct, the tag authenticates the reader, because it knows that only the reader and itself can be aware of the ID and $G_n$. Hence, the first property is satisfied.

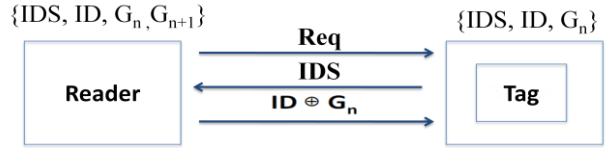

Step 4: The tag calculates two greetings $G_{n+1}$ and $G_{n+2}$ from $G_n$ using its P function, that is,
$$G_{n+1} = P(G_n) \text{ and } G_{n+2} = P(G_{n+1})$$

Then, the tag updates $G_n$ into $G_{n+1}$ and calculates $K_n$ and $K'_n$ from $G_n$ using the public F function:
$$K_n = F(G_n) \text{ and } K'_n = F(K_n)$$

Now, the tag sends $G_{n+1}\oplus K_n$ and $G_{n+2}\oplus K'_n$ back to the reader. Then, the reader can calculate $K_n$ from $G_n$, and recover $G_{n+1}$ from $G_{n+1}\oplus K_n$. If the recovered $G_{n+1}$ is the same with the copy of $G_{n+1}$ it stores, it can authenticate the tag, because it knows that only the tag with the correct function P can generate this $G_{n+1}$. So the second property is also satisfied in our protocol. Once the tag is authenticated, the reader extracts $G_{n+2}$ from $G_{n+2}\oplus K'_n$ using $K'_n$, and updates $G_n$ and $G_{n+1}$ into $G_{n+1}$ and $G_{n+2}$, which will be used in the next round as
$$G_n = G_{n+1} \quad \text{and} \quad G_{n+1} = G_{n+2}$$

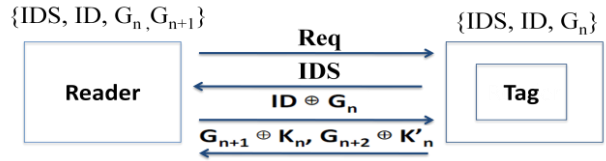

Figure 5. Mutual Authentication

Step 5: Both the reader and the tag update the IDS:
$$IDS_{new} = F(IDS_{old} \oplus G_n)$$

Therefore, an adversary who overhears the old IDS has no way to track the tag.

### 3.2. Ownership Transfer Protocol

Ownership transfer protocol should satisfy the following two properties.

Property 1: The old owner should not be able to access the tag after ownership transfer takes place.

Property 2: The new reader should be able to perform mutual authentication with the tag after the ownership transfer has taken place.

In this case old owner gives the tuple stored for the particular tag (IDS, ID, $G_n$, $G_{n+1}$) to the owner. After one iteration of authentication phase IDS , $G_n$ and $G_{n+1}$ values gets updated and can be known to the only new user.

### 3.3. Linear Feedback Shift Register (LFSR)

A LFSR is a shift register whose input bit is a linear function of its previous state. Mostly used linear function of single bits is XOR, thus normally it is a shift register whose input bit is driven by the exclusive-or (XOR) of some bits of the overall shift register value. LFSR is a simple circuit that consist of shift registers and XOR gates. For an L-bit LFSR, L shift registers are used to store one bit each and can produce $2^L-1$ long sequence of values given an initial non-zero seed. LFSR can be used as pseudo random number generator to obscure transmission between reader and tag.

### 3.4. Physically Unclonable Functions (PUF)

Several different types of Physically Unclonable Functions (PUF)[9] exist. The variant that we will be focusing on is the silicon based PUF (SPUF). SPUFs are delay circuits that take advantage of the fact that no two circuits have exactly the same delay properties, even if they were produced on the same wafer. To produce the random output, a signal is sent through a series of multiplexors switched by input bits. At the end of the circuit there is an arbiter in the form of a flip-flop. In our protocol, PUFs are used to verify the identity of tags. Given a certain input, the tag's PUF will produce a certain output, while other tags' PUFs will produce different output.

Physical Unclonable Functions (PUFs) are innovative primitives to derive secrets from complex physical characteristics of ICs rather than storing the secrets in digital memory. For example, a volatile secret can be generated from the random delay characteristics of wires and transistors. Because the PUF taps into the random variation during an IC fabrication process, the secret is extremely difficult to predict or extract. PUFs significantly increase physical security by generating volatile secrets that only exist in a digital form when a chip is powered on and running. This immediately requires the adversary to mount an attack while the IC is running and using the secret, a significantly harder proposition than discovering non-volatile keys; an invasive attack must accurately measure PUF delays without changing the delays or discover volatile keys in registers without cutting power or tamper-sensing wires that clear out the registers.

In our protocol, PUFs are used to verify the identity of tags. Given a certain input, the tag's PUF will produce a certain output, while other tags' PUFs will produce different output.

## 4. SECURITY ANALYSIS

### 4.1 User Data Confidentiality

The tag ID must be kept secure to guarantee the user's privacy. The tag sends IDS instead of ID. After receiving IDS from the tag, reader sends related entry of ID stored within the reader table using XoR operation with grant number (ID⊕$G_n$).Thus hiding the tag ID to any eavesdropper equipped with an RFID reader.

### 4.2 Tag Anonymity

The location privacy of tag holders can be revealed even when the information on the tag is securely protected. For example, if the messages sent by the tag are well encrypted, the leakage of information is not possible. However, as the tag answers are constant, a tag could be easily associated with its holder. Specifically, location privacy can be more significant when a certain tag is exposed to long-term tracking. It is therefore crucial to make all the information sent by the tag anonymous. This property is guaranteed in all phases of the protocol by generating new IDS in each session.

As seen in section 3.1.2, tags always send fresh IDS, which is obtained by means of a LFSR function which updates IDS during every communication. This property prevents successive tag answers from disclosing any kind of information. Tag anonymity is thus guaranteed and privacy location of the tag owner is not compromised.

### 4.3 Data Integrity

A portion of the tag's memory is rewritable, so modifications are possible. In this part of the memory, the tag stores the index to tag (IDS), Grant number ($G_n$ and $G_{n+1}$). If an attacker does succeed in modifying this part of the memory, the reader will not recognize the tag. As per the property of low-cost tags, they are not tamper resistant. So this kind of tag is susceptible to physical attacks, with the resulting revelation of their content.

### 4.4 Mutual Authentication

We have designed the protocol with both reader-to-tag authentication (ID$\oplus G_n$) and tag-to-reader authentication.

### 4.5 Forward Security

Forward security is the property that guarantees that the security of messages sent today will be valid tomorrow. Since the IDS, $G_n$, $G_{n+1}$ updating is fulfilled after the mutual authentication, a future security compromise on an RFID tag will not reveal data previously transmitted.

### 4.6 Man-in-the-middle Attack Prevention

A man-in-the-middle attack is not possible because our proposal is based on a mutual authentication, in which index to tag (IDS), grant number ($G_n$, $G_{n+1}$) refreshed with each iteration of the protocol, are used.

### 4.7 Replay Attack Prevention

An eavesdropper could store all the messages interchanged between the reader and the tag. Then, he could try to impersonate a reader, re-sending the message (ID$\oplus G_n$) seen in any of the protocol runs. It seems that this could cause the loosing of synchronization between the database and the tag, but this is not the case because after the mutual authentication, the Index to tag (IDS), grant number ($G_n$, $G_{n+1}$) were updated.

### 4.8 Forgery Resistance

The information stored in the tag is sent operated (bitwise XOR ($\oplus$)) with random triplet generated (IDS, $G_n$, $G_{n+1}$) during each iteration. Therefore the simple copy of information of the tag by eavesdropping is not possible.

### 4.9 Data Recovery

The intercepting or blocking of messages is a DoS attack preventing tag identification. As we do not consider that these attacks can be a serious problem for very low-cost RFID tags, our protocol does not particularly focus on providing data recovery.

Table 2 shows a comparison of the security requirements of different proposals in the literature. We have added our proposal in the last column.

Table 2. Security Analysis

| Protocol | SA [12] | PSL [13] | MAP [14] | HB [6] | HB+ [6] | LMAP [5] | ULAP [15] | **proposed** |
|---|---|---|---|---|---|---|---|---|
| User Data Confidentiality | ◯ | ◯ | ◯ | ◯ | ◯ | ◯ | ◯ | ◯ |
| Tag Anonymity | △ | ◯ | ◯ | ◯ | ◯ | ◯ | ◯ | ◯ |
| Data Integrity | △ | △ | ◯ | △ | △ | △ | △ | △ |
| Mutual Authentication | △ | ◯ | ◯ | ◯ | ◯ | ◯ | ◯ | ◯ |
| Forward Security | × | × | ◯ | ◯ | ◯ | ◯ | ◯ | ◯ |
| Man-in-the-middle Attack Prevention | △ | △ | △ | × | × | ◯ | ◯ | ◯ |
| Replay Attack | △ | ◯ | ◯ | ◯ | ◯ | ◯ | ◯ | ◯ |
| Forgery Resistance | ◯ | ◯ | ◯ | △ | ◯ | ◯ | ◯ | ◯ |
| Data Recovery | × | × | ◯ | × | × | × | × | × |

Notation: ◯ Satisfied  △ Partially Satisfied  × Not Satisfied

## 5. EVALUATION

As per the requirement of number of logic gates an arbiter based PUF [9] will require about 8 gates per input bit, plus 4 gates for the arbiter. The inclusion on an LFSR [16] will add an extra 4 gates per bit, plus 3 XOR gates.

LFSR use 4 gates/bit + 3 XOR gates
PUF use 8 gates/bit + 4 gates
For 64 bit key length
LFSR   64*4+3=259 Logic gates
PUF    64*8+4=516 Logic gates
Total =775+9 (control gates) = 784 gates approx

Above mentioned protocol for mutual authentication for low cost tag uses minimal cryptography in terms of logical gates is shown in table 3.

Table 3. Proposed architecture features

| Key Length | | 8-bit | 16-bit | 32-bit | 64-bit | 96-bit | 160-bit |
|---|---|---|---|---|---|---|---|
| Number of Gates | ALU | 103 | 199 | 391 | 775 | 1159 | 1927 |
| | Control | 9 | 9 | 9 | 9 | 9 | 9 |
| | **Total** | **112** | **208** | **400** | **784** | **1168** | **1936** |

## 6. PERFORMANCE ANALYSIS

It is important to carefully analyze the performance of the proposed scheme, to show that it can safely be implemented even in low-cost tags. We will consider the computation overhead, storage overhead and communication overhead for the performance analysis.

### 6.1 Computation Overhead

Low-cost RFID tags are very limited devices, with only a small amount of memory and very constrained computationally. Only 2000 logic gates [10] can be devoted to security-related tasks. As shown in table 3 for 96 bit key also requires 1168 logic gates which is less than threshold value 2000. In our proposal we have taken care that requirement of logic gates should be less than 2000.

### 6.2 Storage Overhead

We assume that all components are L-bit sized, that the IDS size, grant number $G_n$, $G_{n+1}$ values are L-bit. In our case this L can be 8, 16, 32, 64 upto 160 bit. For the implementation of our protocol, each tag should have an memory of size 3L bit as in each iteration we are storing fresh value of IDS and Grant numbers on the tag.

### 6.3 Communication Overhead

The proposed protocol accomplishes mutual authentication between the tag and the reader, requiring only four steps. Taking into account that low-cost tags are passive and that the communication can only be initiated by a reader, four steps may be considered a reasonable number of steps for mutual authentication in RFID environments.

## 7. CONCLUSION

Mutual authentication protocol based on LFSR and PUF function minimizes hardware requirement. Number of gates available in passive tags for security measures are 2000.[10]
As per the table 3 we can use only 784 gates for 64 bit variable and we can extend this up to 160 variable key length. As it requires less hardware, power requirement for processing will be less so we can increase the range of tag detection. If there is attack of message blocking then in that case there are chances to have ambiguity in values of IDS, $G_n$, $G_{n+1}$ of tag and Reader. We can use same protocol with little modification for the tamper detection.

### ACKNOWLEDGEMENTS

I am deeply indebted to my Project Guide Prof. Shankar Nawale whose help, stimulating suggestions and encouragement helped me in all the time of writing of this paper. I want to thank all my colleagues for their help, support, interest and valuable hints.

**Authors**

1. Mr. Amol Bandal pursuing his ME (Computer Engineering) from Pune University. He has 2 yrs industry experience and 3 yrs teaching experience. His area of interest is security and cryptanalysis.

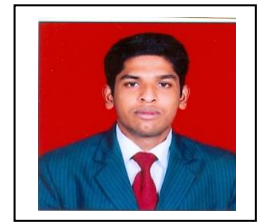